Stefano Bettini[*]

# Anthropic Reasoning in Cosmology: A Historical Perspective

## 1. INTRODUCTION

The so called *anthropic reasoning*[1] is at the centre of an intense debate which in recent times has sparked not only physicists and cosmologists but also philosophers and theologians. One of the main reasons of this wide interest surely resides in the teleological overtones of many recent expositions focused on the topic of the *fine tuning* of some quantities which come out from fundamental physics and cosmology.

The surprising coincidences deriving from dimensionless combinations of fundamental constants of physics and cosmological parameters urged many authors to point out at the special character of many features of *our* universe in innumerable papers and proceedings.[2] They have frequently underlined that the possibility itself of biological complexity depends critically upon the peculiar values of the coupling constants of the four fundamental forces, the slight mass difference between the neutron and the proton, and the actual values of those cosmological quantities that, as in case of the Hubble constant H and the density parameter $\Omega$, govern cosmic evolution.

To quote some well known examples, a change of more than 1% of the strong force coupling constant would result in the absence of elements heavier than lithium; a little modification of the value of the fine structure constant and/or of its gravitational analogue would result in the lack of those main sequence stars which are indispensable for the emergence of complexity and life; a small change in the neutron-to-proton mass ratio would again produce disastrous consequences both for the production of hydrogen itself and the existence of main sequence stars; a modification of the actual values of the above-mentioned parameters H and $\Omega$ (which are connected through Einstein's field equations of general relativity) would result in the production of *universes* inadequate to produce life at some stage of their evolution.

Similar considerations have revealed the existence of a „delicate balance"[3], stimulating three different approaches. The silent majority of physicists, waiting for future developments of their discipline, has assumed a typical "who cares?" attitude, while a consistent part of scholars (including many eminent names of the physics community) found themselves faced with a choice between the following two alternatives:

- 1) the „fine tuning" implies the existence of a „fine tuner"[4]

---

[1] For the meaning of this expression see: Rees, M.J.: *Before the beginning. Our universe and others*, London 1997. Note that Rees does not define explicitly what he means with anthropic reasoning. He simply says that talking of *principles* is „unfortunate", and declares then „to prefer the less pretentious phrase 'anthropic reasoning'". This last expression was used also in the recent Bostrom, N.: *Anthropic bias: observation selection effects in science and philosophy*, London 2002, to summarize a series of topic of anthropic content.

[2] The literature on the topic is immense. The classic texts to become familiar with the arguments involved are: Barrow, J.D.- Tipler, F.J.: *The anthropic cosmological principle*, Oxford 1986; Davies, P.C.W.: *The accidental universe*, Cambridge 1982; Demaret, J – Barbier, C.: Le principe anthropique en cosmologie, *Revue des Questions Scientifiques* 152, 1981, p. 181-222 and 461-509; Demaret, J. – Lambert, D. : *Le principe anthopique, l'homme est-il le centre de l'Univers?*, Paris 1994; Gale, G. : The anthropic principle, *Scientific American* 245, December 1981, p. 154-161; Leslie, J. : *Universes*, London 1989. For extensive bibliographies see Balashov, Y.: Resource Letter AP-1: The anthropic principle, *American Journal of Physics* 59, 1991, p. 1069-1076 and Bettini, S.: *Il labirinto antropico*, available on www.swif.uniba.it/lei/saggi/antropico . It will be re-located to: http://www.swif.it/biblioteca/cxc/ .

[3] Davies, P.C.W.: *The accidental…*, chapter 3.

[4] The expression „divine fine tuner" is used frequently in John Leslie's papers.

- 2) the surprise raised by the „fine tuning" disappears if we consider *our* universe as a very particular member of a large ensemble of "universes" characterized by all possible combinations of initial conditions and fundamental constants.

This dichotomy between a divine Designer and a "many worlds" conception has indeed been at the centre of innumerable discussions and speculations. Moreover, it has also been the main motive of that debate on the anthropic principle which, in the last twenty years or so, has passed from the pages of the physical and astronomical journals to those of popular books and public conferences.

The point is here that the emergence of teleological aspects in the abovementioned discussion came out as a by-product of a technical debate which originally involved a significant group of cosmologists and gravitational researchers and which initially evolved without any kind of teleological overtones (apart few notable exceptions as that due to John Archibald Wheeler[5]).

On the contrary the concept of many different worlds, at least as a counterpart of the idea that our observed region could be not representative of the whole Cosmos, was part and parcel of the anthropic debate since the very beginning.

This history notwithstanding, many people believe that the strict association between anthropic reasoning and teleology was always there. As a matter of fact this belief is largely due to the public debate that followed the publication of Barrow and Tipler's essay *The Anthropic Cosmological Principle* in 1986. Those two authors were surely not the first to perceive or discuss the teleological implications of anthropic principle(s), but their book was surely very influential in promoting a "resurgence of teleological views" if not within the context of contemporary physics, surely in large part of the studies concerning the philosophical meaning of physical researches.

In the following pages I will try to outline a history of anthropic reasoning and principles, limiting myself to the ambit of cosmology, where the arguments emerged. One of my aims will be to show that teleological elements entered the debated topics only at a relatively late , while reflections on many worlds represented a complement of anthropic ideas since the very beginning.[6]

## 2. COSMIC SPECULATIONS AND ANTHROPIC HINTS IN PRE-RELATIVISTIC PHYSICS

To single out an origin for the application of *anthropic reasoning* in cosmology is probably a matter of taste. At any rate, an important preliminary notion for any properly *cosmological* talk consists in a well-defined concept of the universe as a whole.

A non-contradictory quantitative concept of the universe as a self-contained entity was not available until the emergence of general relativistic cosmology in 1917. Before that date, the *universe* was indeed a very vague and undefined concept. As everyone knows, a Newtonian cosmology was in effect beset with unsolvable paradoxes due to the appearance of divergent integrals in all problems containing an infinite and uniformly distributed quantity of matter in an infinite Euclidean space.

The term *universe* itself was not used with great pleasure among scientists if not as a generic name applied exclusively to that *architecture of the heavens* that emerged from observations."[7]

---

[5] Wheeler is a very important, but also a rather *sui generis*, figure in the ambit of contemporary's physics community. For a surely romaned report on his personality see: Overbye, D.: Lo splendente meccanismo dell'universo, *Scienza '81*, 1981, p. 64-70 (and also: Overbye, D.: God's Turnstile: The Work of John Wheeler and Stephen Hawking, *Mercury* 20, n. 4, July/August 1991, p. 98-108.) The use of anthropic reasoning was surely not the first example of a close proximity of Wheeler with teleology. Just to quote an example, think at the implication of the formulation of electrodynamics developed by him and his pupil Richard Feynman in Wheeler, J.A.-Feynman, R.P.: Interaction with the absorber as the mechanism of radiation, *Reviews of Modern Physics* 17, 1945, p. 157-181.

[6] Of course, I am not saying that teleological interpretations of physical results or principles are something new. This will be simply wrong (as showed, for instance, by interminable debates on the meaning of action principles). I am simply saying that the teleological interpretation of the anthropic principle became commonplace mainly after Barrow and Tipler's book.

[7] As significant anecdotes let me say that Clifford affirmed that „in regard to the universe" there was „no right to draw any conclusion at all" (Clifford W.K.: *The first and the last catastrophe. A criticism on some recent speculations about the duration of the universe*, 1874; reprinted in *Lectures and essays*, volume 1, London 1901, p. 222-267, quotation on p. 264), while -three decades later- Rutherford prohibited his students to use that expression during his lessons.

A notable exception was the debate on the global applications of the principles of thermodynamics. Undoubtedly, although not in a rigorous mathematical sense, the term *universe* appeared often in that context in connection with the well-known prediction of a state of thermodynamical equilibrium and mechanical *Heath Death*.

The second *Hauptsatz* posed, amongst other urgent conceptual questions, a major conundrum about the present state of the world; i.e.: why the observed universe appeared so remote from the state of equilibrium?

It is difficult to establish who stated the problem first, but it soon became a major open question. With an eternity of time available it seemed absurd that the *Heath Death* wasn't reached yet, at least if you are not ready to accept a blatant violation of the first principle as that of a special *beginning* of the *universe* as a whole or, at least, of the cosmical surroundings of that particular region of the universe which is inhabited by us.

Just to quote a couple of examples, Vogt discussed the topic around 1878[8], while Fitzgerald, in 1894, asked[9]:

why the ether, the solar system, and the whole universe were not subject to the Boltzmann-Maxwell law?

Amongst the many attempts to furnish answers to questions of this kind, we can distinguish a class of answers of *anthropic* flavour. The first author to suggest something of this kind was presumably Samuel Tolver Preston. This English physicist dedicated large part of his time to thermodynamics, invoking an application of[10] the principles of kinetic theory to the case of the universe not so much as a speculation, but rather as a necessary deduction following from the known principle that detached mass moving freely in space (as the stellar masses are observed to do) and at such distances apart that gravity between the several masses is incompetent to deflect the path of the masses appreciably, must move in straight lines, and have their motions regulated under the mutual encounters in accordance with the principles of the kinetic theory.

In 1879 he suggested that the scale of the universe was „too big" for human observers[11] and that our conclusions were then vitiated by an „extremely limited view".[12] Preston argued that the region of the universe that we inhabit[13] could be atypical because of its „rather exceptional" concentration of hot luminous stars and suggested that[14]:

We may happen to be in a part [of the universe] where the mean temperature of the component matter is exceptionally high, as, of course, from the fact of our being in existence, we must be in a part which is suited to the conditions of life.

Although this passage clearly recalls many contemporary *anthropic* statements, it does not follow that Preston's position was particularly in the vanguard. His treatment was in fact just one among many, in

---

the large debate on the cosmological application of the kinetic theory and on the consequences of energy dissipation. The scenario of the universe depicted by Preston, moreover, was typical of the day and involved, for instance, the presence of a large number of *dark stars* to explain the energetic source of the luminous stars. At last, as for many other authors, the main target of Preston was that of showing how the „stability and permanence" of the „collective universe" could have been assured forever through a „recurring change".[15]

In any case, his *anthropic* solution to the problem of the present state of the observed universe, and the scenario of a large but still local fluctuation in a „boundless universe" globally in the uniform state of maximum entropy, were both original. The author proposed them on several occasions presupposing that the existence of regions, where „the conditions necessary for life" are maintained for periods of time long in comparison to human experience, represented a consistent scenario.[16] Nonetheless Preston was accused of „confusion of reasoning and of unsoundness"[17], forcing him to reply that his arguments implied no violation of the „existing physical principles" in the past and a satisfying „explanation for the existing state of things".[18]

A new version of Preston's scenario was indeed re-proposed less than two decades later by Ludwig Boltzmann. When this happened, the debate on the second *Hauptsatz* and the kinetic theory had already been significantly advanced. During the 1880s, in fact, theoretical physicists had to confront themselves with many experimental inconsistencies and with innumerable difficulties.[19]

In the general debate on these topics particular relevance was reserved to the discussion of Boltzmann's *minimum theorem*. This important mathematical tool represents a relevant piece of the history of the thermodynamics of irreversible processes and has been studied in a historical perspective on various occasions.[20] Here, I'll limit myself to remember that Boltzmann's *minimum theorem*, or *H-theorem* as it was generally known after 1890[21], stated that a particular function describing the behaviour of the inverse of the entropy of a macroscopic isolated system (the *H-function*), must remain, for the greater part of time, at that minimum value which represents the state of maximum entropy.

The status of the theorem remained unclear and raised two kind of controversies: the first, directed basically at the formal aspects of the theorem involved mainly authors of English language and took place essentially after Boltzmann's participation at the *British Association* meeting held in Oxford in August 1894; the second, more radical controversy concerned mainly German authors and was directed against the whole statistical interpretation of the second *Hauptsatz* and the probabilistic nature of the H-theorem.

The British controversy consisted in an exchange of letters and short contributions which appeared on the pages of *Nature* in the years 1894/1895. It was in this context that Fitzgerald question about the present state of the universe, originally raised at the Oxford meeting, was re-proposed by both Bryan and Culverwell.[22]

---

In the February 28th, 1895 issue of *Nature* Boltzmann clarified his position and argued that the H-theorem demanded simply „that in the course of time the universe must tend to a state where the average *vis viva* of every atom is the same", but evidently found that this was not enough to evade Fitzgerald's objection. Surely he was worried by a conception which presumed a very special arrangement of the whole universe at a certain time (if not a very peculiar beginning), finding hard to reconcile it with the mechanical *Weltbild*.

As a way out he took refuge to an idea attributed to his „old assistant" Ignaz Schütz[23], that was expressed in the following terms[24]:

We assume that the whole universe is, and rests forever, in thermal equilibrium. The probability that one (only one) part of the universe is in a certain state, is the smaller the farther this state is from thermal equilibrium; but this probability is greater, the greater the universe itself is. If we assume the universe great enough we can make the probability of one relatively small part being in a given state (however far from the state of thermal equilibrium), as great as we please. We can also make the probability great that, though the whole universe is so far from thermal equilibrium, our world is in its present state. It may be said that the world is so far from thermal equilibrium that we cannot imagine the improbability of such a state. but can we imagine, on the other side, how small a part of the whole universe this world is? Assuming the universe great enough, the probability that such a small part of it as our world should be in its present state, is no longer small.

If this assumption were correct, our world would return more and more to thermal equilibrium; but because the whole universe is so great, it might be probable that at some future time some other world might deviate as far from thermal equilibrium as our world does at present. Then the aforementioned H-curve would form a representation of what takes place in the universe. The summits of the curve would represent the worlds where visible motion and life exist.

This argument was repeated by Boltzmann at least two times in the following years.[25] On these new occasions the author no longer credited Schütz, and presented the idea of fluctuations of cosmical proportions as his personal opinion. This presumably was a consequence of the polemical exchange with Ernst Zermelo which had started in December 1895.

It is well known that, contrary to the English authors that participated in the *Nature* debate, Zermelo didn't try simply to eliminate the contradictions between the mechanistic foundations of the H-theorem and the irreversibility of entropy. He rather affirmed the absolute value of the second *Hauptsatz*, rejecting any mechanistic interpretation.

The specific points of the Zermelo/Boltzmann controversy have been told many times[26], but what is relevant here is the answer of Boltzmann to certain „questions of principle" [*principielle Fragen*] advanced by his rival.[27]

During the debate, it was made evident that a non-contradictory statistical explanation required the adoption of an „unverifiable assumption" [*unbeweisbare Annahme*] according to which the universe, or at least a very large part of it, is at present in an „improbable state" because it „began from a very improbable state".[28] Boltzmann found this conclusion, as everyone based on „special conceptions on the universe" [*specielle Vorstellungen über das Universum*], very unpleasant.[29] According to him, in fact, there was another conception which was consistent with the mechanical representation of the world, i.e.: that, suggested in the 1895 paper, of a universe globally in thermal equilibrium containing here and there „relatively small ambits" [*verhältnissmässig kleine Bezirke*] (called *single worlds* [*Einzelwelten*]) and imagined of „the extension of our stellar space" [*der Ausdehnung unseres Sternenraums*]) which can be far from that state.[30]

Such a scenario, apart from evading the unsatisfactory picture of an unilateral change of the whole universe from a determined initial state to a final terminal state", was considered by Boltzmann as the best *image* available of the "world as a mechanical system".[31]

### 3.ANTHROPIC SUGGESTIONS IN THE ERA OF THE EXPANDING UNIVERSE

The fundamental characteristics of a (general relativistic) mathematical model of the whole universe became clear thanks to Einstein's theory of gravitation and to the association between the revolutionary idea of an expanding universe and Hubble's empirical law. General relativistic cosmology became indeed centred on a particular class of space-times of constant curvature: those which are spatially homogeneous and globally isotropic or, in other words, obey the cosmological principle.

As a result, after the general acceptance of the idea of an expansion of space in 1930, many reviews were published to point out the dynamical behaviour of the many possible homogeneous and isotropic universes (the Fridman-Lemaitre-Robertson-Walker or FLRW *universes*) which are regulated by that simplified version of Einstein's equations known as Fridman's equations.[32]

The (Preston)/Schütz/Boltzmann argument raised various critical reflections before 1930[33] and was then sometime discussed in the new context of the expanding universe. Some authors considered it „completely false"[34] because of the implausibility of a fluctuation as large as the region of space visible to Mount Wilson reflector[35]; others simply found it incomprehensible.[36] Someone as influential as the British biologist John Burdon Sanderson Haldane, attributed a certain relevance to Boltzmann's hypothesis arguing that „in the course of eternity any event with finite probability will occur".[37]

Tolman recalled Boltzmann's argument on 1931 in one of his enquiries on the meaning of the entropy of the universe in the context of a relativistic thermodynamics.[38] There he remembered the old difficulty of classical thermodynamics expressed by Fitzgerald's question, confiding to have learnt Boltzmann's answer through Tatiana Ehrenfest.[39] Aiming at the foundation of a cogent relativistic thermodynamics, Tolman found all possible classical way outs, including Boltzmann's, wrong or incomplete answers to the above-mentioned question. Anyway, he depicted the „fluctuations theory" as an „important possibility" and an „important part" of any future „relatively complete treatment of the entropy of the universe".[40]

Coming from one of the most *empirically* oriented cosmologists[41], this appreciation of the fluctuation hypothesis cannot be a coincidence. Boltzmann's scenario, transformed by Tolman into that of an inhomogeneous universe with no temporal beginning, represented indeed an alternative to those simple models that, as FLRW universes, were extrapolating an *a priori* assumption of spatial homogeneity beyond the observable region.

Tolman considered FLRW universes as extraordinarily important geometrical tools in virtue of their mathematical simplicity, but he was always clear that they were only rough idealizations not to be confounded with the actual physical universe. In 1934 he even proposed to study the „effects of inhomogeneity on the theoretical behaviour of cosmological models", underlying the necessity of being not „too dogmatic" about accepting conclusions deducted from FLRW models.[42]

---

[33] E.g.: Nabl, J.: Der zweite Hauptsatz der Thermodynamik und der Satz von der Entropie im Lichte des Boltzmannschen H-Theorems der Gastheorie, *Naturwissenschaftliche Rundschau* 21, 1906, p. 337-341; Borel, E.: *L'espace et le temps*, Paris 1923; Weyl, H.: *Philosophie der Mathematik und Naturwissenschaft,* München 1927.

[34] Bronstein M.P.-Landau L: Über den zweiten Wärmesatz und die Zusammenhangsverhältnisse der Welt in Großen, *Physikalische Zeitschrift der Sowjetunion* 4, 1933, p. 114-118, on p. 117.

[35] Cf. Bronstein M.P.: On the expanding universe, *Physikalische Zeitschrift der Sowjetunion* 3, 1933, p. 73-82, in particular p. 74.

[36] E.g.: Takeuchi, T.: On the cyclic universe, *Proceedings of the Physico-Mathematical Society of Japan* (3) 13, 1931, p. 166-177, in particular p. 166.

[37] Haldane J.B.S.: The universe and irreversibility, *Nature* 122, 1928, p. 808-809, on p. 809. I consider here anyway the post-1930 Haldane's book edition *The inequality of man and other essays*, London 1932. The author calculated that an improbable distribution as that requested by Boltzmann's thesis was so improbable to demand something as $10^{10^{100}}$ years to happen. Notwithstanding he concluded that -in a truly cosmological perspective- the hypothesis of fluctuations represented the most reliable explanation of the apparent contradiction between the observed state of the universe and the predictions of thermodynamics.

[38] For a general review of Tolman's ideas see: Tolman R.C.: Thermodynamics and relativity, *Bulletin of the American Mathematical Society*. 39, 1933, p. 49-74.

[39] Tolman, R.C.: On the entropy of the universe as a whole, *Physical Review* 37, p. 1639-1660, 1931, on p. 1642. Note that T. Eherenfest came back on Boltzmann's argument in the preface of the English edition of Ehrenfest, P. & T.: *Begriffliche… (cf. n. 16 above)*. See*: The Conceptual Foundations of the Statistical Approach in Mechanics*, Ithaca (NY) 1959, p. xi.

[40] Tolman, R.C.: On the entropy…, p. 1660.

[41] On the topic see: Merleau-Ponty, J.: *Cosmologie du XX siècle*, Paris 1965; Eisenstaedt J.: Cosmology: a space for thought on general relativity, in: *Foundation of big bang cosmology. Proceedings of the seminar on the foundations of big bang cosmology*, Meyerstein W.F. (ed.), Singapore, 1989, p. 271-295.

[42] Tolman, R.C.: Effects of inhomogeneity on cosmological models, *Proceedings of the National Academy of Sciences of the United States of America* 20, 1934, p. 169-176, on p. 176. Tolman suggested also the possibility of a universe of variable curvature with open and closed regions, pointing out at a scenario that was developed both in Soviet *cosmological* speculations and in various *many universes* proposals of the 1960ies.

Such an attitude echoed, probably without Tolman's knowledge, that already suggested - not without a certain polemical vein towards Einstein's static cosmology - by Emile Borel in the 1920ies.[43] This French author transformed Boltzmann's argument in a warning according to which „local knowledge cannot give knowledge of the universe".[44] He then compared our conditions as Earth-based-observers to that of a fictitious inhabitant of a drop of water who remains unable to see the complexity beyond his/her abode.[45]

Anyway, apart from the reservations against the cosmological principle in its deductive formulation and the exploration of a model-universe inhomogeneous on a very large scale[46], Tolman underlined also the *anthropic* content of Boltzmann's hypothesis. He suggested in fact implicitly that the existence of „sentient beings" represented by itself a valid argument against the „enormous improbability" of the requested fluctuation.[47]

These topics (criticism of cosmological deductivism, proposal of an inhomogeneous scenario and *anthropic* justification of the special characteristics of our observable region) were all revived in the mid 1950ies in a paper which appeared on the *Newsletter* of the astrophysical Institute of Kazakhstan and that was later considered by Ya.B. Zel'dovich as the first application of the *anthropic principle* in relativistic cosmology.[48]

The paper in question[49], called *Essential features of the astrophysical observed universe as typical properties of the inhabited cosmic system*, was written in Russian by Grigory Moiseevich Idlis and was published in 1958 although the author remembers to have developed his ideas two years before.[50]

Of course, we must consider Idlis' paper at the light of both the status of the worldwide cosmological controversy that characterized the 1950ies and of the peculiar situation of the whole discipline of relativistic cosmology in the USSR a few years after the death of Stalin. As practically any paper published there, Idlis's article revealed a mutual contamination between the concepts of dialectical materialism and those emerging from observational and theoretical cosmology.[51] Moreover, it shared with all the other Soviet papers devoted in some way to cosmology, the picture of a universe that was infinite[52], eternal and inhomogeneous on a very large scale.[53]

---

Idlis anyway accepted the general relativistic interpretation of the red-shifts, limiting himself to raise objections against the linearity of the velocity-distance relation. As usual he referred to the observed expanding part of the universe as the *metagalaxy*[54], suggesting to consider it a very peculiar region of the infinite universe; i.e.: a particular system of galaxies, extended over at least five billion light years and approaching in first approximation an isotropic and homogeneous cosmological model with characteristic age, average density, average temperature and expansion rate.

The heart of Idlis' monography consisted in connecting the „characteristic features" of the observed region of the universe to the properties necessary for the rise, evolution and maintenance of life.[55]

Idlis considered, firstly, the conditions necessary for the emergence of life on a local astronomical scale (a typical work in what today we would call astrobiology, in the spirit of many classic papers of Idlis' mentor: Vasilii Grigor'evich Fesenkov[56]). In the second part of his paper the Soviet astronomer then discussed the properties necessary for the emergence of life in the large structure of the universe, pointing out the peculiarity of an approach which was looking for a „coherent solution" of the properties of the observed region from the „fact itself of our existence".[57]

At last, he concluded that any typical inhabited system would have shared with our observable expanding *metagalaxy* the fact of necessarily being an „isolated" region of the universe endowed with an appropriate age, density, temperature and chemical composition.

Following Boltzmann, he stated moreover that there was no reason for appealing to „anomalous initial conditions" of the whole universe. In fact, maintaining that living beings can observe only regions of the universe that do possess the properties of a typical habitated system and not a region whatever in the „infinite multiformity" of the universe, we have no right to extrapolate the properties of our observable region to the whole.[58]

In his rehabilitation of Boltzmann's hypothesis, Idlis rejected the objections expressed in name of the improbability of a fluctuation of cosmical proportion by Soviet authors as Bronstein, Landau and Zel'manov.[59] He remarked that the existence of a „thinking being" does not simply demand the „appearance of a habitable solar system", but rather implies as a rule the generation of a very large quantity of uninhabitable planetary systems, stars and galaxies. Therefore, in conclusion, Idlis suggested that the realization of fluctuations „initially devoid of structure" on a metagalactic scale was not so unlikely and presumably „necessary for the appearance of the living beings that observe the picture of the world extending before us".[60]

---

universe scenario which was appreciated in the Soviet Union as a way out of Olbers' paradox in the context of an infinite universe. Cf. Idlis, G.M.: Теория относительности и структурная бесконечность Вселенной [The theory of relativity and the structural infinity of the universe], *Astronomicheskii Zhurnal* 33, 1956, p. 622-626.

[54] The term *metagalaxy* concerns here the expanding system of galaxies in the observable part of the universe. It was adopted by Lundmark and Shapley in the 1920ies to avoid any ambiguous expression regarding an unobservable whole. Later it was used by different critics of FLRW cosmology (as, for instance, the followers of Klein-Alfven cosmology) and also in the papers of astronomers not inclined to cosmological speculations (just to quote an example, Vera Rubin adopted it in her celebrated Rubin V.C.: Differential rotation of the inner metagalaxy, *Astronomical Journal* 56, 1951, p. 47).

[55] Life was seen here as a „regular exit of matter evolution". Idlis (on p. 55 of his 1958 paper) adopted Oparin's interpretation of the definition of life given in Engels' *Dialectic of nature*. In such a perspective life represents a special form of the movement of matter or, in other words, an emerging quality that comes out from the movement of matter at a certain stage. Cf. Oparin, A.I.- Fesenkov, V.G.: *Life in the universe*, New York 1961 (Russian original 1956).

[56] Fesenkov and Idlis were very close to each other. In the late 1930ies Fesenkov suffered as many other astronomers persecutions that comported his dismissal from the position of chairman of the Astronomical Council in 1937. In 1941 - during second world war - Fesenkov was then extradited to Kazakhstan's capital, Almaty. There he was able to organize a new institute of astronomy. Idlis became the new director of the Institute after Fesenkov moved to Moscow. Later, Idlis moved to Moscow too, and began to work for the History of Science's Institute.

[57] Cf p. 39 of Idlis' 1958 paper.

[58] Ibid., p. 52

[59] Apart from his criticism against Boltzmann's argument, Abram Leonidovich Zel'manov made some interesting *anthropic* considerations in his writings. He suggested in fact not only a *predisposition* of the properties of the metagalaxy respect to the development of life, but even the existence of „qualitatively different" areas of the *metauniverse*. He argued moreover that other universes (if they exist at all) could be doomed to evolve without witnesses. Contrary to Idlis, Zel'manov did not formulate his ideas quantitatively. This was probably due to the fact that, because of his Hebrew origins, he found it very hard to publish in astronomical journals (still being able, however, to exert a certain influence on his colleagues).

[60] All the quotations in the last paragraphs are from p. 53 of Idlis' 1958 paper.

Idlis' ideas came out almost simultaneously with the papers of Gerald James Whitrow and Robert Henry Dicke; two authors very far from Idlis' materialistic attitude that advanced *anthropic* answers to some old problems, yet sustaining at the same time evolutionary FLRW cosmology.

Whitrow, in particular, proposed an *anthropic* resolution of the venerable philosophical question *Why physical space has three dimensions?*[61] (arguing that with a space of different dimensionality there would be no living being to pose the question) and, similarly to Idlis, alluded around 1955 to an *anthropic* explanation of the size of the observable universe. Anyway he never published these last ideas, which were developed years later by Wheeler.[62] The only reference to Whitrow's argument that appeared in print during the 1950ies seems to be that due to the philosopher of religion Eric Lionel Mascall, who attributed to the English's mathematician that[63] it may be necessary for the universe to have the enormous size and complexity which modern astronomy has revealed, in order for the earth to be a possible habitation for living beings.

## 4.THE EPOCH OF MAN

The decade that began in 1953 saw a *renaissance*[64] of gravitational research. Controversy between the general relativistic evolutionary cosmology and the steady-state one was „most intense".[65] Among the basic events of that period one may count:
-   Rindler's paper on horizons[66];
-   Ryle's results on the distribution of radiosources (which testified the first strong observational evidence against steady-state cosmology[67]);
-   the theory of stellar nucleosynthesis developed by Hoyle, Fowler and the Burbridges, that was accompanied by Hoyle's *anthropic prediction* of a $^{12}C$ resonance level around 7,68 MeV[68] (a

---

[61] Whitrow, G.J.: Why physical space has three dimensions?, *British Journal for the Philosophy of Science* 6, 1955, p. 13-31 and *The structure and evolution of the universe*, second edition, New York 1959. For a critical discussion of Whitrow's arguments see for instance: Smart, J.J.C. 1987: Philosophical problems of cosmology, *Revue International de Philosophie* 41, p. 112-116; Leslie, J.: Anthropic principle, world ensemble, design, *American Philosophical Quarterly* 19, 1982, p. 141-151. An antecedent to Whitrow's arguments might be found in: Ehrenfest, P. 1917: In what way does it become manifest in the fundamental laws of physics that space has three dimensions?, *Koninklijke Akademie van Wetenschappen te Amsterdam. Section of Sciences, Proceedings* 20, 1917, p. 200-209; Welche Rolle spielt die Dreidimensionalität des Raumes in den Grundgesetzen der Physik?, *Annalen der Physik* 61, 1920, p. 440-446.

[62] E.g.: Wheeler, J.A.: The universe as home for man, *American Scientist* 62, 1974, p. 683-691; The beam and stay of the Taub universe, in: *Essays in General Relativity: a Festschrift for Abraham H. Taub*, Tipler F.J. (ed.), New York 1980, p. 59-70. Wheeler ideas, in turn, were criticised in Shepley, L.C.: Tidal forces in a highly asymmetric Taub universe, in: *Essays in General Relativity…*, p. 71-77. Cf. also Barrow, J.D.–Tipler, F. J.: *The Anthropic …*, ch. 6.3.

[63] Mascall, E.L.: *Christian theology and natural science*, London 1956, p. 43. Someone has noted that the link between the size of the universe and the presence of life within it was touched by Edgar Allan Poe in his poem in prose of 1848 *Eureka*. Cf. Cappi, A.: Edgar Allan Poe's physical cosmology, *Quarterly Journal of the Royal Astronomical Society* 35, p. 177-192.

[64] Cf. Pais, A.: *'Subtle is the Lord …'. The science and life of Albert Einstein*, Oxford 1982, ch. 15.

[65] Kragh, H.: *Cosmology and controversy. The historical development of two theories of the universe*, Princeton 1996, p. 392.

[66] Rindler, W.: Visual horizons in world-models, *Monthly Notices of the Royal Astronomical Society* 116, 1956, p. 662-677.

[67] E.g.: Ryle, M.-Scheuer, P.A.G.: The spatial distribution and the nature of radio stars, *Proceedings of the Royal Society* 230, 1955, p. 448-462; Ryle, M. - Clarke, R.W.: An examination of the steady-state model in the light of some recent observations of radio sources, *Monthly Notices of the Royal Astronomical Society* 122, 1961, p. 349-362; Hewish, A.: Extrapolation of the number-flux density relation of radio stars by Scheuer's statistical method, *Monthly Notices of the Royal Astronomical Society* 123, 1961, p. 167-181.

[68] Hoyle's prediction of this particular resonance level of carbon-12 has been often regarded as an *anthropic* one. In the course of his enquiries on stellar nucleosynthesis, Hoyle considered the reaction that bring to the formation of $^{12}C$ from three $^{4}He$ nuclei, and deduced that it depends crucially on the existence of an energy level of $^{12}C$ (the 7,656 one) which is just above the rest mass of a $^{8}Be$ nucleus and a $^{4}He$ nucleus. He was then amazed by the consequences of this coincidence. In fact, if it was not for that particular resonance level, carbon would be extremely rare in the universe. But it was not all: Hoyle noted moreover that $^{16}O$ presented a peculiar resonance level too (the 7,1616 MeV level). If this last energy level was just a little higher, almost all the $^{12}C$ would have been turned in oxygen.

In the Mid 1960ies (Hoyle, F.: *Galaxies, nuclei and quasars*, London 1965, ch. VI), Hoyle argued that these coincidences could be typical of our part of the universe and suggested that in other *portions* of the universe, the resonance levels could be result that was indeed presented already in 1953 after a „private communication" of Hoyle himself[69]);

- a certain debate on Gamow's theory of creation, followed by all the puzzles and the suspicious remarks raised by the conception of a radical violation of physical laws and energy conservation in an initial singularity.

The dispute between the rival theories of cosmology was not the only open front. Of considerable interest to cosmologists of any school was the old problem of the coincidences between large dimensionless numbers, obtained through the combination of some fundamental quantities of physics and cosmology. This problem was crucial already for Weyl[70], few years after the formulation of general relativity, and became largely known thanks to Eddington and Dirac.[71]

Motivated by the peculiar coincidence in order of magnitude between the age of the universe expressed in atomic units[72], $H_0^{-1}[(m_p c 2\pi)/h]$ (which is at present a pure number of order $10^{39}$), and the large pure number of order $10^{39}$ given by $e^2/(Gm_p m_e)$, Dirac advanced in the 1930ies a *new basis for cosmology*. This was based on the „fundamental principle" (which was later renamed the *large number hypothesis*, or LNH for short) according to which all dimensionless numbers of order $(10^{39})^n$ must vary proportionally to the n-th power of the age of the universe expressed in atomic units.

As a particular consequence of this cosmological principle we have that the number given by $e^2/(Gm_p m_e)$ - or, analogously, what we today call the reciprocal of the dimensionless gravitational fine structure constant, $\alpha_G^{-1} \approx (hc)/(2\pi G m_p^2)$ - should vary with the age of the universe. It follows then that at least one of the presumed constants must vary, changing approximately[73] as a simple function of the age of the universe.

placed differently, with the result of having no living creatures around. He consequently presented a cosmological scenario according to which „the universe would be far richer in its possibilities and content than we normally imagine. In other regions the numbers would be different and the gross properties of matter, the science of chemistry for example, would be entirely changed" (Hoyle, F.: Recent developments in cosmology, *Nature* 208, 1965, p. 111-114, on p. 114).

Of course, Hoyle preferred a similar scenario to the eventuality of some sort of finality in nature. Still worried by teleology he successively rejected the *anthropic principle* as a sort of blind alley. Anyway, he confessed at least once to be against the support offered by the *anthropic principle* to the *big bang religion*, rather than against the *strong anthropic principle* in itself (Hoyle, F.: The Anthropic and perfect cosmological principles/similarities and differences, unpublished manuscript of the talk presented at the *second Venice conference on cosmology and philosophy* on November 1988). For some recent analysis of Hoyle's *anthropic* argument, see: Livio, M.- Hollowell, D.-Weiss, A.-Truran, J.W.: The anthropic significance of the existence of an excited state of 12C, *Nature* 340, 1989, p. 281-284; Jeltema, T.E.-Sher, M.: The triple-alpha process and the anthropically allowed values of the weak scale, *Physical Review* D61, 2000, p. 017301; Oberhummer, H.-Pichler, R.-Csoto, A.: The triple-alpha process and its anthropic significance, nucl-th/9810057 v2, 1999; Oberhummer, H.-Csoto, A.-Schlatt, H.: Stellar production rates of carbon and its abundance in the universe, *Science* 289, 2000, p. 88-94. For Hoyle's personal recollections, see: Hoyle, F.: *Home is where the wind blows. Chapters from a cosmologist's life*, Oxford 1997, ch..XVIII.

[69] Dunbar, R.E.P.-Pixley, R.E.-Wenzel, W.A.-Whaling, W.: The 7.68 MeV state of C[12], *Physical Review* 92, 1953, p. 649-650

[70] Weyl, H.: Eine neue Erweiterung der Relativitätstheorie, Annalen der Physik (4) 59, 1919, p. 101-133. On Weyl's contributions see: Bettini, S.: *Dalla cabala dei grandi numeri ai principi antropici*, degree thesis in philosophy, Firenze 1990; Gorelik, G.: Hermann Weyl and large numbers in relativistic cosmology, in: *Einstein studies in Russia. Einstein studies, vol. 10*, Balashov, Y.-Vizgin, V. (eds.), Boston 2002, p. 91-106.

[71] On the topic cf. among others: Harrison, E.R.: The cosmic numbers, *Physics Today* 25 (12),1972, p. 30-34; Wessson, P.S: *Cosmology and geophysics*, Bristol 1978; Barrow, J.D.: The lore of large numbers: some historical background to the anthropic principle, *Quarterly Journal of the Royal Astronomical Society* 22, 1981, p. 388-420; The mysterious lore of large numbers, in: *Modern cosmology in retrospect*, Bergia, S.-Balbinot, R.-Bertotti, B. (eds.), Cambridge 1990, p. 67-93; Kragh, H.: Cosmo-physics in the thirties: towards a history of Dirac cosmology, *Historical Studies in the Physical Sciences* 13, 1982, p. 69-108; Cosmonumerology and empiricism: the Dirac-Gamow dialogue, *Astronomical Quarterly* 8, 1991, p. 109-126; Bettini, S.: *Dalla cabala…*; Kilmister, C.W.: *Eddington's search for a fundamental theory. A key to the universe*, Cambridge 1994.

[72] For instance: $(e^2/m_e c^3) \approx 10^{-23}$ or $h/m_p c^2 = 0.46$ $[e^2/(m_e c^3)]$. In the formulae given here and in the text $H_0$ is the present value of the Hubble constant; $h/2\pi$ is the Dirac's form of Planck's constant (that I'm writing here explicitly); e is the electron charge; G is Newton's gravitational constant; $m_p$ and $m_e$, are respectively the mass of the proton and of the electron.

[73] i.e.: apart for small numerical coefficients.

Dirac suggested that there was one changing *constant* and that this was G. This launched a debate beyond the narrower domain of theoretical physics which became particularly intense during the 1950ies.[74]

Of particular interest was of course Dicke's treatment, which emerged from a deep analysis of the foundations of general relativity and of any gravitational theory which could represent a reliable alternative to Einstein's.

Dicke also invented the microwave radiometer and described the cosmic background radiation discovered by Penzias and Wilson in 1964 as relic of the *primeval fireball*. At about 1955 he dedicated himself to the development of experimental tests of gravity, evaluating the existing proofs in favour of general relativity and submitting to a severe analysis Einstein's principle of equivalence.

The eventuality of a varying G surely intrigued him (as the Brans/Dicke theory testifies), but at the same time he surely didn't share Dirac's rationalistic approach to mathematical physics with his invocation to „elegance, simplicity and perfection".[75]

As to the peculiar coincidence between $\alpha_G^{-1}$ and the present age of the universe, Dicke concluded that there was no need at all to invoke a variability of G. To explain that coincidence it was in fact sufficient to re-consider the statistical premises of Dirac's reasoning, without indulging in conclusions based on aesthetic criteria. In short, while for Dirac the present epoch has obtained completely fortuitous; according to Dicke the age „now" is „not random" but rather „conditioned" by „biological factors"[76], because we can say in advance that the existence of observers (i.e.: carbon based life forms) is allowed only on a limited temporal range of the evolutionary history of the universe.

Dicke stated that we (and whatever chemically complex alien civilization) may observe the universe only in those epochs that present the necessary conditions for our own existence since 1957, but the clearest (although probably the less known) illustration of his argument was advanced in occasion of the *Joseph Henry Lecture* held in front of the *Philosophical Society of Washington* on April 18, 1958. There he affirmed[77]:

To infer the time dependence of the gravitational interaction requires more than a simple observation that the reciprocal of the gravitational constant and the age of the universe, when expressed dimensionlessly, are *now* nearly equal. It is also necessary to assume that *now* is a random time. But is it?

The present epoch is conditioned by the fact that the biological conditions for the existence of man must be satisfied. This requires the existence of a planetary system and a hot star. If we assume an evolutionary cosmology starting with the formation of hydrogen 12 billion years ago, there is an upper limit for the epoch of man which is imposed by the following two conditions: First, hydrogen is being continually converted to helium and heavier elements. Perhaps 20% has already been "burned." Second, there is an upper limit on the radiating life of a star.

If the star is massive (10 times the sun's mass) it lives riotously, burning its hydrogen like a wastrel. For a light star (1/10 the sun's mass), hydrogen is burned slowly and the star is capable of living much longer than the sun,

---

[74] Among the innumerable contributions to a question which posed more interrogatives than solutions, Jordan proposed various cosmological models inspired by Dirac's LNH; Bondi discussed widely the topic of large numbers coincidences in his classic textbook *Cosmology*; Hoyle affronted the theme in many occasions; Oskar Klein tried to elaborate a solution in the context of a cosmological model completely different from both evolutionary and steady-state theories. See the sources quoted on n. 65 above for further details.

[75] Dicke, R.H.: Gravitation without a principle of equivalence, *Reviews of Modern Physics* 29, 1957, p. 363-376, on p. 363.

[76] Ibid., p. 375. A similar argument was almost incidentally delineated by Eddington in his *Messengers Lectures* of 1934. Discussing the eventuality of „a fortuitous deviation of entropy from its maximum value", he said: „the year 1934 is not a random date between t=-∞ and t=+∞. We must not argue that because fluctuations of the present magnitude occupy only 1/xth of the time between t=-∞ and t=+∞, therefore the chances are x to 1 against such a fluctuation existing in the year 1934. For our present purpose the important characteristic of the year 1934 is that it is selected as belonging to a period during which there exist in the universe beings capable of speculating about the universe and its fluctuations. It is clear that such creatures could not exist near thermodynamical equilibrium. Therefore it is perfectly fair for the supporters of this suggestion to wipe out of the calculation all those multillions of years during which the fluctuations are less than the minimum required to permit of the evolution and the existence of mathematical physicists". See Eddington, A.S.: *New pathways in science*, Cambridge 1935, p. 65.

[77] Dicke, R.H.: Gravitation-an enigma, *American Scientist* 47, 1959, p. 25-40, on p. 33 (Dicke's italics). The paper appeared originally in 1958 on the *Journal of the Washington Academy of Sciences*, vol. 48.

100 times as long. However, if the star is much smaller than this, its central temperature never rises high enough to cause nuclear reactions to take place. Such a light star radiates until its gravitational energy is gone and then it cools off. It is seen therefore that the longest life of a star is very roughly $10^{14}$ years and this puts an upper limit to the epoch of man.

There is also a lower limit on the epoch of man. With the assumption that initially only hydrogen exists, it is necessary to produce other elements in the stellar caldrons and distribute them about the universe before a planetary system of our type can be formed. It is a bit difficult to estimate this time, but it would seem that 1 billion years would be a reasonable lower bound on the epoch of man. It is thus seen that the epoch of man is not random but is very roughly delineated.

The consecration of the abovementioned argument came only after the publication of a letter by Dicke in the November 4, 1961 issue of *Nature*. That letter, where the Princeton's physicist gave to his arguments a definite mathematical formulation, is now very famous among scholars, especially for that passage stating that the Hubble age of the universe „is not a 'random choice' from a wide range of possible choices, but is limited by the criteria for the existence of physicists".[78]

Dirac replied immediately[79] and – although admitting not to have a „decisive argument" against Dicke's „assumption" – wrote in favour of his LNH because it enabled an indefinite existence of habitable planets and the consequent possibility of a never ending life.[80]

Preferring a theory that admits an eternal presence of life in the universe could appear curious, but it surely wasn't a new argument. For instance, Tolman called the eventual reduction of humanity to a „transitory and improbable phenomenon" an „emotional" (rather than intellectually founded) objection to Boltzmann's hypothesis[81], while Sciama confessed around 1960 that one of the „most important" issue in favour of the steady-state was that of being „the only model in which it seems evident that life will continue somewhere".[82]

Dicke, on his part, argued in retrospect to have suggested a „very conservative statement" to a „rather straightforward question".[83] Also, when his argument became known as an application of the *weak anthropic principle*, he affirmed that there was nothing of particularly „exciting" in it when confronted with the idea of a „natural selection of the natural constants" that was contemplated some years later by Brandon Carter.[84]

## 5. BECOMING AWARE OF THE DELICATE BALANCE

The 1960ies saw the gradual affirmation of the standard general relativistic hot big bang model as the result of various factors. Apart from the already mentioned evidences deriving from the surveys of distant radiosources and the accumulation of other astrophysical evidence (concerning quasars, X-ray sources, etc.) the two basic *proofs* that corroborated the model were the theoretical capacity of predicting the observed abundances of light atomic nuclei and the discovery of the microwave background radiation. Together with (or: as part of) the newly emerging paradigm, the period was characterized by an increasing rapprochement between cosmology and particle physics and by the formulation of the essential mathematical background needed for the study of the causal structure of space-time and the meaning of singularities.

---

[78] Dicke, R.H.: Dirac's cosmology and Mach's principle, *Nature* 192, 1961, p. 440-441. Quotation from p. 440.

[79] Dirac's reply follows immediately after Dicke's letter on *Nature* 192, 1961, p. 441.

[80] It seems pretty absurd that the hopes of Dirac, that Carter in recent times pitilessly called an „error of blatant wishful thinking" (Carter, B.: The anthropic principle: self-selection as an adjunct to natural selection, in: *Cosmic perspectives*, Biswas, S.K.–Malik, D.C.V. – Vishveshwara, C.V. -eds.-, Cambridge 1988, p. 183-204, on p. 188), were lately made an *anthropic principle* by Tipler (i.e.: the so called *Final Anthropic Principle*). This FAP has said to be based on the most beautiful of all physical postulates: „total death is not unavoidable". Cf. for instance Tipler, F.J.: The Omega Point theory: A model of an evolving God, in: *Physics, philosophy and theology: A common quest for understanding*, Russell, R.J. –Stoeger, W. R.-Coyne, G. V. (eds.), Vatican Observatory 1988, p. 313-331; *The physics of immortality. Modern cosmology, god and the resurrection of the dead*, New York 1994; Barrow, J.D.-Tipler, F.J.: *The anthropic cosmological ...*, ch. 1 and 10. For a historical discussion see chapter 11 of Bettini, S.: *Il labirinto antropico ...*, and references quoted therein. For a recent analysis of Tipler's ambitions cf Goenner, H.: *The quest for ultimate explanation in physics: reductionism, unity, and meaning*, Max-Planck-Institut für Wissenschaftsgeschichte, preprint 187, Berlin 2001.

[81] Tolman, R.C.: On the entropy..., p. 1642.

[82] Quoted on Kragh, H.: *Cosmology and...*, p. 254 (taken from an interview to Sciama of April 14, 1978 belonging to the American Institute of Physics' collection. See: www.aip.org ).

[83] R.H. Dicke in Lightman, A.-Brawer, R.: *Origins, the lives and worlds of modern cosmologists*, Harvard 1990, p. 210-211. Cf. also Dicke's opinions as referred in Pagels, H.R.: A cozy cosmology, *Sciences* 25, 1985, p. 34-38.

[84] Cf. Misner, C.W.-Thorne, K.-Wheeler, J.A.: *Gravitation*, San Francisco 1973, p. 1217.

A particularly important centre for the development of new ideas, methods and techniques was surely Cambridge's DAMTP (*Department of Applied Mathematics and Theoretical Physics*) where a new generation of researchers (including Hawking, Rees and Ellis) developed new ideas under the supervision of Dennis Sciama. Because of the presence of Hoyle and the inheritance of people like Eddington and Dirac, Cambridge already represented an ideal place for cosmological studies.[85] Sciama's *relativity group* gathered some of the best minds assuring them „a certain structure and someone who was willing to take them on".[86]

Carter was one of Sciama's students: born in Sydney in 1942, he arrived at DAMTP around 1964 and got his Ph.D. in applied mathematics and theoretical physics in 1968. The largest part of his published papers of the 1960ies and early 1970ies were dedicated to the Kerr solutions of Einstein's field equations (i.e: those exact solutions describe the field of rotating black holes) or to other technical problems related with the global properties of exact solutions in general relativity. In these fields Carter obtained relevant results, including a theorem that now bears his name[87] and the demonstration of the existence of a family of charged Kerr solutions.[88]

In 1967 Carter drew up a long type-written preprint devoted to *the role of fundamental microphysical parameters in cosmogony* which remained unpublished.[89] This was conceived as the first part of a projected work (called *the significance of numerical coincidences in nature*) aimed to furnish an „unified treatment" and a „readily accessible and comprehensible" survey of numerical coincidences emerging from physics and astrophysics. In short, the paper represented a stimulating exercise of what Victor Weisskopf would call *qualitative physics*[90]: a genre of physical discussion which had some forerunners (from Galilei's reflections on the size of the animals attributed to Salviati in the second day of his *Discorsi e dimostrazioni matematiche*[91] to Edwin E. Salpeter's speculations of the mid 1960ies)[92] and a long list of successors[93] and that, in Carter's words, concerned „the manner in which familiar local

---

phenomena depend qualitatively, and in order of magnitude, quantitatively on the fundamental parameters of microphysics".[94]

The author explored in particular the role and consequences of a series of „relations and coincidences" obtained through the composition of the following six fundamental parameters:

-   the pseudo-scalar coupling constant of strong interactions $g_s$,[95] which is approximately 4 in the so called *Planck units*; i.e. those fundamental units such that $c = G = h/2\pi = 1$[96]
-   the electron charge e ($\approx 1/12$ in the above-mentioned units)[97]
-   the nucleon mass $m_N$ ($\approx \frac{1}{2}$ X $10^{-19}$)[98]
-   the ratio between the pion ($\pi$-meson) mass and the nucleon mass, $m_\pi/m_N \approx 1/7$, which suggests the „maximum effective range of the strong interactions"
-   the ratio between the mass of the electron and that of the nucleon: $m_e/m_N \approx 1/1830$
-   the ratio between the difference between neutron and proton mass and the mass of the nucleon, $\Delta_N/m_N \approx 1/730$.

From the interplay of these parameters it is possible to show not only that the sizes and masses of the planets and the stars must lie in certain typical ranges, but that they have a fundamental role in determining „the character of all important natural phenomena" with the exception of those where high energy physics or cosmological quantities are directly involved. Carter showed how (apart from „relatively small adjustment factors") most of the „limiting masses of astrophysics arise (in fundamental units) simply as the reciprocal of the gravitational fine structure constant". There was only one notable exception, and to point this out represented the most significant result reached in the 1967 unpublished paper.

This exception concerns the positioning of the dividing line which distinguishes main sequence stars (i.e.: stable hydrogen's burning stars) that transport energy mainly by convection (i.e.: red dwarfs) from those in which energy is dissipated mainly by radiative transport (i.e.: blue giants).[99] Carter showed that this line „occurs within the range of main sequence stars only as a consequence of the rather exotic coincidence that the ninth power of the electromagnetic fine structure constant [i.e.: $e^2$ in Carter's notation] is roughly equal to the square root of the gravitational fine structure constant"[100]. This coincidence implied indeed a delicate balance. „Had it been the 11$^{th}$ power", he wrote, „all main sequence stars would be convective red dwarfs". This observation, which prompted some crucial developments, will be re-considered below.

Notice that the connection between „most of the large numbers of cosmogony" and powers of $\alpha_G^{-1}$ was known since the 1930ies. The merit of Carter lies not only in gathering a series of results, but also in explicitly warning against „misconceptions" as those elaborated by Pascual Jordan.[101] Jordan, in fact,

---

16

conjectured an „unconventional" cosmological mechanism (involving the age of the universe as in Dirac's LNH) to give an explanation of that large-number coincidence which relates the upper limit of the number of nucleons contained in a star and $\alpha_G^{-3/2}$.

Although this coincidence was a genuine finding, there was no need to postulate any new physics to explain it. Jordan's coincidence is in fact *predictable* in terms of the ordinary theory of stellar evolution, taking into account that the masses of stable stars are necessarily related to the Landau/Chandrasekhar limiting mass.[102]  It is enough to realize that „no normal stable star can exist with a nucleon number which differs from the Landau number [$N_L \equiv M_L/M_N \equiv M_N^{-3}$ in Carter's units] by more than a factor of order 10²".[103]

In other words we cannot have the formation of a star with $M << M_L$, because if that were the case gravity would be balanced by quantum mechanical pressure due to the exclusion principle as it actually happens in the case of planets. At the same time we cannot have the formation of a star with $M > 10^2 M_L$ because such an object would be very unstable due to the predominance of radiation pressure.

## 6. FROM COGNIZABILITY TO THE ANTHROPIC PRINCIPLE

Brandon Carter himself does not remember today what happened to the second part of his 1967 paper.[104] We know anyway that his „ultimate purpose" was there „to clarify the significance" of the coincidence between $\alpha_G^{-1}$ and $H_0^{-1}[(m_p c 2\pi)/h]$.[105]

According to Carter, that coincidence should have been „fully explained in principle (although many relevant details remain uncalculable in practice) in terms of conventional physics and cosmology" without the recourse to „revolutionary departures" such as the LNH or Eddington's *Fundamental Theory*. Programmatically he then stated that the „final task" of the 1967 preprint was to furnish the formulae which illustrate the „connection between local and cosmological quantities … via the timescales of stellar evolution" in terms of microphysical parameters.

If not the second part of that paper, the promised follow-up came out three years later in the form of a new 13-page-typescript for the *Clifford Memorial Meeting* held at Princeton University on February 21, 1970.  On that occasion, with Wheeler and Dyson in the audience, Carter expressed the purpose to „clarify" the „various much publicised" large number coincidences „in terms of standard physical theory in conjunction with the orthodox hot big bang model of the universe".[106]

He classified the „coincidences and interrelations" amongst the large adimensional numbers of „cosmogonical" interest in three distinct categories:

- The first one included coincidences whose explanation was accounted for completely in terms of "objective certainties" (i.e.: without requiring any recourse to „statistical probabilities") without leaving current physics and astrophysics. Jordan's coincidence represented an example of this category.

- „Category (II)" contemplated „those coincidences whose explanation, although straightforward, requires subjective and probabilistic considerations relating to our own position as observers in the universe". It included, of course, the Dirac/Dicke coincidence.[107]

- To category (III) belonged:

those coincidences which cannot be given a direct physical explanation since they depend more or less critically on the actual values of fundamental or microphysical constants, but which nevertheless could in principle have been predicted in advance of their observational discovery on the ground that they are necessary preconditions for the existence of observers (ourselves) in the universe.

As example of this kind of coincidences Carter invoked the relation elaborated by Sciama on 1953 in his analysis on an inertial induction law of Mach's type[108]:

$$G\rho_0 H_0^{-2} \approx 1$$

According to Carter, this expression[109] „could have been predicted from the conventional idea that the density irregularities in the form of galaxies, stars etc, which are presumably necessary for our own existence, grew from initially small perturbations of the homogeneous background".

Once accepted the „conventional" theory according to which galaxies form by condensation, starting as small density fluctuations of the homogeneous FLRW background, the task became to make evident the biological constraints that the fact itself of our existence (being the growing of initial density irregularities, and hence the formation of galaxies, a necessary pre-condition for the emergence of life), imposes not simply on our temporal location in the universe but on some fundamental characteristics of the universe itself. In this case, Carter firstly choose two „independent constants" which are responsible of the dynamics of the FLRW background (and consequently of the temporal evolution of parameters such as the Hubble constant H, the curvature K, the average density $\rho$ and the black body temperature T):

- the ratio between the baryon number density (which is a conserved quantity) and the third power of the cosmological black body temperature in a certain $\eta \equiv n_b/T^3$ (which „represents roughly the ratio of the mean non relativistic gas pressure to the electromagnetic black body radiation")

- the curvature scalar of the homogeneous 3-space sections at constant cosmic time[110] $\zeta \equiv K/T^2$

He then pointed out at the fact that, according to the Fridman equations, we may have Sciama's coincidence only if curvature does not dominate on matter density at present time, and argued at last the biological constraints on $\zeta$ deriving from the need that density irregularities transformed effectively in galaxies at some moment of the evolutionary history of the universe.

---

[107] In Carter's notation it is expressed as $H(t_0) \approx m_N^3$. Following Dicke, Carter discussed here the constraints imposed by the existence of observers to the „observed value of the cosmological time", suggesting some further cosmological insights and founding a reason to reject by principle not only Dirac's „revolutionary departure from orthodox physical theory" but also any form of steady-state theory that postulates an independence between the age of the universe and the Hubble constant.

[108] Sciama, D.: On the origin of inertia, *Monthly Notices of the Royal Astronomical Society* 113, 1953, p. 34-42. Cf. also: Sciama, D.: *The unity of the universe*, London 1959. As for the case of Eddington, Dirac and Jordan, Carter learned of Sciama's coincidence by Bondi's treatment in *Cosmology* (Bondi, H.: *Cosmology*, second edition, Cambridge 1961). Carter affirmed then, in 1973, that an anthropic explanation would not have „ruled out here the possibility (or desiderability)" of a Machian framework underlying ordinary gravitational theory. See: Carter, B.: Large number coincidences and the anthropic principle in cosmology, in: *Confrontation of cosmological theories with observational data*, Longair, M.S. (ed.), Dordrecht 1974, p. 291-298, on p. 295.

[109] That, in Carter's notation, results $\rho_0 \approx H_0^{-2}$ (being G = 1)

[110] A similar discussion is available in Carter, B.: Large number coincidences…, p. 293/295, where the attention is concentrated on the total lifetime of closed universes which come out from the radiation era and are then regulated by the dominant contribution of matter density. The implicit preference for the closed case was supported by observational data practically until the publication of Gott III J.R.-Gunn, J. E.-Schramm D.N.-Tinsley, B.M.: An unbound universe?, *Astrophysical Journal* 194, 1974, p. 543-553. Since then the majority of cosmologists began to prefer the open universe in the face of observational evidence, although sometimes invoking the flat case for theoretical reasons (as those suggested by many inflationary models).

To guarantee the growth of density irregularities, Carter firstly underlined that one must allow for the decoupling of matter from radiation pressure; an event that requires that T drops „well below"[111] the Rydberg ionization energy.[112] Additionally, he noticed that one must have the curvature K not too different in order of magnitude from the density at the time of decoupling. In fact, if K had had „a too strongly negative value" the kinetic energy of expansion would have dominated the potential energy, making impossible the re-contraction of the perturbations under the action of gravity and causing their dispersal together with the expansion of the universe. If, to the contrary, K had had a too strongly positive value it would have provoked an early contraction of the whole universe, with the consequent destruction of the developing condensations.

In the light of these two considerations Carter concluded that our existence (or if you want: the possibility of galaxy formation) imposes the inequality (in Planck's units)

$$|\zeta| << 10^{-2}e^4m_e(\eta m_N + 10^{-2}e^4m_e)$$

in order to restrict the present value of K so to account for Sciama's coincidence within the observational accuracy available.

He also suggested that the coincidence between the number of baryons in the visible universe and the square of $\alpha_G^{-1}$, which had become famous through Eddington's papers, appeared as a consequence of Sciama's relation once the present age of the universe was accounted for.[113]

Apart from these rather technical details a crucial aspect was, at any rate, Carter's awareness that it was hard to ascribe the status of a proper physical explanation to the arguments exploited to predict Sciama's relation. Physicists were used to extend the available knowledge in order to derive parameters previously taken as fundamental from „something more basic", but this surely wasn't the case here.

It is at this point that the concept of an „ensemble of universes" appeared for the first time, invoked as an essential element to promote the kind of reasoning illustrated above in the case of the growth of density irregularities – which is often labelled as an anthropic prediction – to the status of *explanation*. Carter wrote[114]:

However it is worth bearing in mind that all category III predictions can be raised automatically to the status of genuine explanations if we are willing to adopt some sort of statistical world-ensemble philosophy. In this type of philosophy one postulates the existence of an ensemble of universes characterized by all possible combinations of initial conditions and fundamental constants (the distinction between these concepts, which is not at all clear cut, being that the former refer to essentially local and the latter to essentially global features) with some probability measure, the assignment of which presents a deep problem. The measure problem can however be by-passed to a considerable extent, because the existence of any organism describable as an observer will only be conceivable for certain restricted combinations of the fundamental constants, which distinguish within the world-ensemble an exceptional cognizable subset, to which our own universe must necessarily belong. (more detailed, but for practical purposes unfeasible, consideration of the detailed local conditions would distinguish within the cognizable subset a cognate subset in which observers actually occur.) A category III prediction is equivalent to a demonstration that the features under consideration is common to all members of the cognisable subset, which thus explains why it is present in our own universe.

With the introduction of this *world-ensemble* philosophy Carter attempted to define no less than a new goal to physical inquiry. He described it as the aim to show that[115]:

---

[111] „well below" means here „several powers of ten", Cf. Carter, B.: Large number coincidences…, p. 294.

[112] Which is ½ e⁴mₑ in Planck's units. Cf. Misner, C.W.-Thorne, K.-Wheeler, J.A.: *Gravitation*, …, chapter 28.

[113] Of course this not furnish yet an answer on the particular order of magnitude of $\alpha_G^{-1}$. Carter will look for „a possible explanation" of that large number recurring to a world-ensemble philosophy. His treatment of Eddington's coincidence was extended in Carter, B.: Large number coincidences…, p. 294/295.

[114] I'll follow here the original paper in adopting underscorings.

[115] Carter noted anyway that the aim to reduce „all the main global constants from fundamental to derived status" in a „completely satisfactory" manner was largely illusory „because of the lack of a hard and fast distinction between global and local parameters". The same distinction between initial conditions and fundamental constants was then unclear because the first appealed to essentially local and the latter to essentially global characteristics.

the cognizable subset is so closely circumscribed that the principal global constants, while remaining genuinely fundamental, are nevertheless forced to have values very close to those actually observed.

In practice this appears as a „very hard" task, but what is relevant here is above all a matter of principle. Once accepted as a premise that some characteristics (as, for instance, the fact of being composed of chemical elements produced in stellar interiors or the need for habitable planets steadily heated by stars) must be common prerequisites for the existence of all the possible forms of intelligent life in the universe, we could concentrate ourselves on finding „a fairly complete system of restrictions on the values of fundamental constants". In other words we could think to collect all the „necessary restrictions" that the presence of biological complexity imposes on the values of fundamental constants in any possible *cognizable* universe.

The peculiar examples advanced by Carter exploited of course many of the relations discussed in 1967.[116] He showed for instance that if the coupling constant of the strong interactions $g_S^2$ was only a little weaker, there would be only hydrogen around; while if it had had a little larger value probably we would have „stable nuclei of an almost unlimited size".[117] Particular attention was given to the „remarkable coincidence" that governs the subdivision between red dwarfs and blue giants.[118] This time, Carter underlined how critical the order of magnitude of $\alpha_G^{-1}$ (rather than that of $\alpha^{-1}$) was and advanced a „potential category III explanation" of the weakness of that coupling constant. He suggested that[119]:

the formation of a planetary system may be dependent on the passage of a star through a Hayashi convective phase shortly before reaching the main sequence. Since planetary formation theory is not yet on a sound footing this idea is of course entirely speculative, but empirical evidence in its favour is provided by the observation that red dwarfs, which are still convective and middle sized stars like the sun, which left the Hayashi phase only just before reaching the main sequences, have much lower angular momenta than blue giants – the implication being that the former but not the latter lost angular momentum in forming planets. If this idea is correct, then a universe in which the gravitational coupling is significantly stronger than the value given by [$e^{20} \approx m_N$] would not merely lack convective stars on or near the main sequence, but would in consequence have no planets and therefore no people.

Speculative as it is[120], this argument testifies above all the attempt to render acceptable the category III predictions as explanations.

---

[116] In particular the following four coincidences: $g_S^2 \approx 2m_N/m_e$; $\Delta_n/m_e \approx 2$; $\alpha \approx \Delta_n/m_e$; $g_S \approx (1/3)\alpha^{1/2}$. Barrow and Tipler (Barrow, J.D.-Tipler, F.J.: *The anthropic cosmological* …, p. 400 and note 48, p. 452) have emphasized in particular the relevance of the coincidence $\Delta_n/m_e \approx 2$, which – once written in the form $\Delta_n - m_e \approx m_e$, as in Carter's 1967 preprint, results „crucial for the existence of a live-supporting environment in the universe" (in connection with the productions of the adequate percentage of light elements in the early phases of the cosmological evolutionary history). Barrow and Tipler state here that their „belief that Carter's [1967] work should appear in print provided the original motivation for writing" their celebrated essay.

[117] On these points Cf also Carr, B.J.-Rees, M.J.: The anthropic principle …., p. 611 and Barrow, J.D.-Tipler, F.J.: *The anthropic cosmological* …, p. 398-400.

[118] Carter reports here the coincidence as $e^{20} \approx m_N$, adding that it is „satisfied empirically within about a factor of ten". In front of the fact that the value of $m_N$ is around 8 X $10^{-20}$, an approximation of this kind is probably more appropriate of that reported in 1967. Anyway it is a frequent practice, in questions related to orders of magnitude, to make approximations in the range of two orders of magnitude. This practice was for instance justified by Dirac in his early papers on LNH. See in particular: Dirac, P.A.M.: A new basis for cosmology, *Proceedings of the Royal Society A165*, 1938, p. 199-208.

[119] Cf. the analogous passage in Carter, B.: Large number coincidences…, p. 297. Note that Carter's argument avoids here the eventuality of having a value of $\alpha_G^{-1}$ smaller than the actual one. To the topic is dedicated just a short remark at the end of his Princeton's talk. There Carter suggests that the problem probably implies „a specific assumption about the fundamental state vector" in an Hilbert-space context, in order to favour "moderate rather than extreme values of the basic coupling constants". In this perspective $\alpha_G^{-1}$ „would be explained as the least extreme value compatible with the existence of observers" producing a new kind of explanation that should be classified as *category IV*.

[120] See the remarks on Barrow, J.D.-Tipler, F.J.: *The anthropic cosmological* …, p. 336.

As noted by Carter in the concluding part of the 1970 talk, all depends on „one's attitude to the world-ensemble concept"[121] or, to better say, on the acceptance of „the idea that there exist many universes, of which only one can be known to us". Although „philosophically objectionable" at first sight, a similar conception recurred on various occasions in the 1960ies and, to be honest, also before.

Of course we should now clarify the meaning of the plural *universes*, but I can only hint at this very delicate question here. In short, if the word *universe* means (as it does) an all-inclusive physical whole, every recourse to the plural form should simply imply a misuse of language.[122] In spite of this fact cosmologists have anyway used that plural since the very early days of relativistic cosmology[123], not only with reference to different theoretical world-models but also intending causally disjointed regions within a certain model (as, in particular, in the context of the so called Eddington-Lemaitre model, a particular type of the FLRW class of models[124] where expansion is accelerated by the presence of a positive cosmological constant and „disconnected universes"[125] are generated every time that „neither light nor any other causal influence" will be able to pass from one region to another[126]).

However, if we agree on terminology we could talk properly of different *universes* also in the case of separated regions of a universe that is inhomogeneous on a very large scale as the one depicted by Idlis.

In the 1960ies talk on many universes became relatively common when, motivated by different aims, various authors extended their speculations well beyond the Eddington-Lemaitre or any other of the FLRW models. As an example let me recall here the attempt (due to Fred Hoyle and Jayant V. Narlikar) of saving the concept of global stationarity through a „radical departure" from  the original steady state model.[127] Hoyle and Narlikar postulated that, on a very large scale, there were many „individual regions" possessing different physical properties; similar scenarios were presented – almost in the same period – by Jaroslav Pachner and Ronald Gordon Giovanelli.[128]

What matters here is that the concept of *causally separated* regions in a larger (if not infinite) space-time, found many different representations before 1970. Carter, on his part, added something to the Boltzmann /Idlis scenario; to wit, the idea that different *universes* may exhibit peculiar values of the constants of nature. He was not simply imagining regions of the universe where parameters as the temperature or the mean density may assume different values, but rather an ensemble of universes each regulated by a different physical phenomenology.

---

To say it clearly, Carter was describing a metaspace of universes whose constants were free to assume all possible values. We must note, anyway, that he denied as „unconventional" the temporal variation of the fundamental constants in any single member of the metaspace, maintaining that in each particular universe the peculiar values of the constants, once fixed, was maintained for the whole of its evolutionary history.

A conception of *co-existing* universes of this kind[129] was presumably derived from three distinct sources. Firstly, from the lectures held in Cambridge in the early 1960ies by biologist Charles Pantin (that Rees remembers to have attended together with Carter[130]); secondly, from the „unconventional" scenario of separated *bubbles* described by Hoyle and Narlikar around 1966[131]; thirdly, and mainly, from the *relative state formulation* of quantum mechanics due to Hugh Everett which became called *many worlds interpretation* after de Witt contributions.[132]

It is in fact the de Witt-Wheeler interpretation of Everett's proposal, with its description of a universe[133] possessing „many branches, only one of which can be known to any well defined individual observer, but all of which are equally real", that Carter invoked in his 1970 paper in association with the „statistical ensemble" requested by Category III explanations.

„Everett philosophy" – as Carter baptised it – not only appeared as „the only interpretation of quantum theory … which makes sense in cosmological contexts", but also as the appropriate and natural tool for considering some of the coupling constants as particular operators in the Hilbert-space of a world-ensemble.

# 7. THE ANTHROPIC PRINCIPLE(S)

---

[129] The expression co-existing universes is mediated from a classification due to George Gale (see: Gale, G.: Cosmological fecundity …). He distinguished three classes of many universes scenarios: „spatially multiple universes", „temporally multiple universes" and „other-dimensional multiple universes". The third class is defined as those universes which exist simultaneously as single spatiotemporal entities in some sort of metaspace (Gale does not say this clearly, but it is implicit in his exposition which is centred on the concept of possibility). Ambiguous as it is, Gale's classification is at bottom nothing more that the re-elaboration of the Medieval distinction among the innumerable speculations on many worlds which followed Etienne Temper's decree of 1277. On this last point cf. in particular Duhem, P.M.M.: *Medieval cosmology: theories of infinity, place, time, void, and the plurality of worlds*, Chicago 1986.

[130] Rees, M.J.: *Before the beginning …*, p. 259. Pantin suggested in particular that „if we could know that our own Universe was only one of an indefinite number with varying properties we could perhaps invoke a solution analogous to the principle of Natural Selection; that only in certain universes, which happen to include ours, are the conditions suitable for the existence of life, and unless that condition is fulfilled there will be no observers to note the fact." Cf Pantin, C.F.A.: Life and the conditions of existence, in: *Biology and Personality*, Ramsey, I.T. (ed.), Oxford 1965, p. 83-106, on p. 103-104.

[131] Carter alluded to a peculiar aspect of Hoyle and Narlikar work in Carter, B.: The complete analytic extension of the Reissner-Nordström metric in the special case $e^2 = m^2$, *Physics Letters* 21, 1966, p. 423-424. He didn't show, anyway, any clear sign of being interested at all in their new cosmological ideas.

[132] I will spend just few words on Everett's relative state formulation of quantum mechanics here. In his original papers of the late 1950ies the author presented a conception in which measurement didn't collapse the wave function to a single value. Everett didn't write a lot on the topic and, apart from the purpose of saving a realistic interpretation of quantum mechanics, it remains largely unclear how this theory should work in order to give determinate measurements results.
During the 1960ies, however, Everett's formulation was associated with a many-worlds interpretation. This was primarily due to the reading of Everett by Wheeler and Bryce de Witt. The latter, in particular, suggested explicitly that the world was splitting in many alternative real branches as a result of quantum measurements.
De Witt published his views at last on the pages of the September 1970 issue of *Physics Today* (provoking an extended debate, which was covered in the April 2001 issue of that journal) and then practically canonized them in an anthology of papers on Everett's proposal – edited jointly with Graham – that was called *The Many-Worlds Interpretation of Quantum Mechanics*, a name that was there to stay.
The MWI was destined to become a proposal seen with suspicion by almost everyone with the exception of cosmologists, which found in it the only way to connect the concept of a wave function of the whole universe with the embarrassing eventuality to conceive an observer external to the universe itself.

[133] Or to better say: of „the state vector of the universe"

Five/six years after the discovery of the cosmic microwave background the hot big bang model was on the verge to become *standard* and to enter in textbooks as a paradigm.[134] The consecration of the model, which consisted essentially in the physical description of the evolution of the early expanding universe from $10^{-2}$ seconds after the big bang to the decoupling at the end of the radiation era, was in fact canonized early in the 1970ies in a long series of books and papers.[135]

Clearly the model was not (and still is not) a complete theory. Just to give some examples, the formation of cosmic structures, the peculiar features of the hadronic era or of those which preceded it, or the initial singularity itself all remained problems unsolved in the context of the model itself and waiting for further physical advancements.

In effect is hard to say today what Carter exactly meant with *conventional* or *orthodox* physics. If *orthodox* stands for *accepted* knowledge in a certain moment, we must not forget that many aspects of the soon-to-come *standard* model of elementary particle physics were still part of a work in progress in the early 1970ies. At the same time, the emerging paradigm produced new *enigmas*, among them the first discussions of the so called *flatness* and *horizon* problems.[136]

These problems contributed to transform the fears of an oversimplification of FLRW cosmologies that pestered the researchers of the 1930ies into a series of interrogatives on the peculiar features of the actual universe. In particular, the isotropy of the microwave background became a fact in need of a coherent explanation rather than an ideal (and presumably oversimplified) assumption.

In the second part of the 1960ies cosmologists began to follow two main schools: some invoked a mechanism capable to smooth the universe content in the very early evolutionary phases[137]; while others took the observed situation at its face-value posing questions on the peculiarity of initial conditions.

The *Dicke/Carter philosophy*, as it was soon called[138], found a collocation in this second school when it was used by Collins and Hawking to establish how peculiar were the initial conditions that generate the observed spatially isotropic universe with respect to the set of initial conditions of all possible spatially homogeneous universes that emerge as solutions of Einstein's field equations (with a minimum of physical assumptions on the energy contents).

In their paper of 1973 the two English physicists regarded indeed Carter's ideas on a very large number of universes as the „most attractive answer"[139] to the puzzling situation that our highly isotropic

---

[134] Take care of the fact that almost everything cosmologists use the term paradigm in a sense very different from Kuhn. In fact, they generally mean with "paradigm" a network of new ideas or theoretical options which integrate (rather than substitute through a "scientific revolution" in Kuhn's sense) the preceding ones. Moreover, sometimes cosmologists use the term "paradigm" (instead of "model") to mean a theoretical structure which is not yet completely formalised and which is consequently still open to personal beliefs. To check some examples cf. Ellis, G.F.R.: Innovation, resistance and change: the transition to the expanding universe, in: *Modern cosmology in retrospect*, Bertotti, B.-Balbinot, R.-Bergia, S.-Messina, A. (eds.), Cambridge 1990, p. 97-114; Kolb, E.W.-Turner, M. S.: *The early universe*, Redwood City, California 1990 (in particular p. 313-314); Coles, P.-Lucchin, F.: *Cosmology. The origin and evolution of cosmic structure*, Chichester 1995 (in particular p. xii).

[135] E.g.: Peebles, P.J.E.: *Physical cosmology*, Princeton 1971; Weinberg, S.: *Gravitation and cosmology*, New York 1972; Harrison, E.R.: Standard model of the early universe, *Annual Review of Astronomy and Astrophysics* 11, 1973, p. 155-183; Misner, C.W.-Thorne, K.-Wheeler, J.A.: *Gravitation*, … .

[136] Dicke was probably the first to note in an evident way the flatness problem on p. 62 of Dicke, R.H.: *Gravitation and the universe. The Jayne Lectures for 1969*, Philadelphia 1970. That paper anticipated indeed of a decade or so the famous Dicke, R.H.-Peebles, J.; The big bang cosmology-Enigmas and nostrums, in: *General relativity: an Einstein centenary survey*, Hawking, S.W.-Israel, W. (eds.), Cambridge 1979, p. 504-517.

[137] After: Misner, C.W.: Transport processes in the primordial fireball, *Nature* 214, 1967, p. 40-41; Neutrino viscosity and the isotropy of primordial blackbody radiation, *Physical Review Letters* 19, 1967, p. 533-535; The isotropy of the universe, *Astrophysical Journal* 151, 1968, p. 431-457.

[138] By both Wheeler and Collins and Hawking.

[139] Collins, C.B.-Hawking, S.W.: Why is the universe isotropic?, *Astrophysical Journal* 180, 1973, p. 317-334, on p. 334. A first draft of the paper was received by the *Astrophysical Journal* on June 29, 1972. A revised version followed on September 25, 1972.

universe was of „measure zero in the space of all homogeneous models".[140] Consequently they concluded[141]:

From the existence of the unstable anisotropic mode it follows that nearly all of the universes become highly anisotropic. However these universes would not be expected to contain galaxies, since condensations can grow only in universes in which the rate of expansion is just sufficient to avoid recollapse. The existence of galaxies would seem to be a necessary precondition for the development of any form of intelligent life. Thus there will be life only in those universes which tend toward isotropy at large times. The fact that we have observed the universe to be isotropic is therefore only a consequence of our own existence.

The Collins/Hawking paper was not the only one that invoked Carter's *line of thought* before it first appeared in a publication in 1974. Freeman Dyson (who was probably the first to appeal in 1972 at something as *Carter's principle of cognizability* on print) esteemed Carter's „speculative arguments" and his extension of Dicke's ideas[142]; Martin Rees expounded Carter ideas and terminology on a couple of occasions, arguing amongst other things that Dicke's coincidence had to be „automatically satisfied" in any *cognizable* universe.[143] Also Tryon was presumably influenced by Carter's ideas (although he didn't quote him) in introducing a „principle of biological selection, which states that any universe in which sentient beings find themselves is necessarily hospitable to sentient beings" in his paper on the creation of the universe out of nothing.[144]

A strong support for Carter's perspectives then came from John Archibald Wheeler, who linked the world-ensemble concept with his cyclical model, where new closed universes with different values of various fundamental parameters emerged after the Big Crunch of a previous universe.[145] Already in December 1967, when discussing the concept of superspace, Wheeler contemplated the „thought-provoking" ideas of his Princeton colleague Dicke, when addressing the question as to „why then do we happen to be living in that part of superspace where we find ourselves?".[146] In the early 1970ies he quoted several times the Dicke/Carter's *philosophy* and Carter's still unpublished arguments.[147] In particular, at the symposium *on the development of the physicist's conception of nature* held in Trieste on September 1972, Wheeler had the following verbal exchange with Dirac[148]:

---

[140] Ibid., p. 333. The assumption of a *young* universe implies an alternative to the conclusion according to which it is of zero measure in the considered metaspace. This eventuality has been studied by Barrow in different occasions. E.g.: Barrow, J.D.: The isotropy of the universe, *Quarterly Journal of the Royal Astronomical Society* 23, 1982, p. 344-357; Barrow, J.D.-Sonoda, D.H.: Stability of certain spatially homogeneous cosmological model, *General Relativity and Gravitation* 17, 1985, p. 409-415; Asymptotic stability of Bianchi type universes, *Physics Reports* 139, 1986, p. 1-49 Barrow, J.D.-Tipler, F.J.: *The anthropic cosmological ...*, section 6.11.

[141] Ibid., p. 319. A statement often quoted from the Collins and Hawking's paper was: „the answer to the question "why is the universe isotropic?" is "because we are here" " (cf p. 334).

[142] Dyson, F.J.: The fundamental constants and their time variation, in: *Aspects of quantum theory*, Salam, A.-Wigner, E.P. (eds.), Cambridge 1972, p. 213-236, on p. 235.

[143] Rees, M.J. 1972: Cosmological significance of $e^2/Gm^2$ and related large numbers, *Comments on Astrophysics and Space Physics* 4, 1972, p. 179-185, on p. 181. See also Rees' paper on *The far future* in John, L. (ed.) 1973, *Cosmology now*, London 1973. In the latter paper Rees touched the topics of an ensemble of universes and of a natural selection of the natural constants, but he named only Wheeler and not Carter.

[144] Tryon, E.P.: Is the universe a vacuum fluctuation?, *Nature* 246, 1973, p. 396-397. Other authors discussed Carter's *cognizability*, but their papers were published in 1974 or later. Among these are Harrison (who, arguing an „anthropomorfic conception of intelligence", criticized the metaphysical nature of the cognizability's principle in Harrison, E.R.: Cosmological principles II. Physical principles, *Comments on Astrophysics* 6, 1974, p. 29-35, on p. 30-31) and Ellis (Ellis, G.F.R.: Cosmology and verifiability ..., on p. 259).

[145] E.g.: Misner, C.W.-Thorne, K.-Wheeler, J.A.: *Gravitation*, ..., section 44.6; Patton, L.M.–Wheeler, J.A.: Is physics legislated by cosmogony?, in: *Quantum gravity: an Oxford symposium*, Isham, C.J.-Penrose, R.-Sciama, D. W. (eds.), Oxford 1975, p. 538-605; Wheeler, J.A: Genesis and observership, in: *Foundational problems in the special sciences*, Butts, R.E.-Hintikka, J. (eds.), Dordrecht 1977, p. 3-33.

[146] Wheeler, J.A: Our universe: the known and the unknown, *American Scientist* 56, 1968, p. 1-20, which is an adaptation of a communication presented on December 29, 1967 for New York's *American Association for the Advancement of Science*.

[147] E.g: Misner, C.W.-Thorne, K.-Wheeler, J.A.: *Gravitation*, ..., p. 1216-1217.

[148] During the discussion that follows Dirac, P.A.M.: Fundamental constants and their development in time, in: *The physicist's conception of nature*, Mehra, J. (ed.), Dordrecht 1973, p. 44-54. Quotations are from p. 58 of Mehra's volume.

J.A. Wheeler: How do you feel about the explanation of Brandon Carter that many cycles of the universe are possible and the constants in this particular cycle are such as will permit life?
P.A.M. Dirac: That doesn't get over the difficulty that you have to explain this very big number

In the end Wheeler asked Carter „to say something for the record" on his ideas. In other words: to publish something on a subject by now widely discussed in the community of astrophysicists and cosmologists but that is author, even considering it „potentially fertile", still believed in need of „further development".[149]

The opportunity arrived in occasion of the 63th symposium of the IAU, held in Cracow from 10th to 12th September 1973. That meeting, dedicated to the *Confrontation of Cosmological Theories with Observation*, coincided with the celebrations of the 500th anniversary of Copernicus' birth and Carter spoke in the section devoted to the *structure of singularities*. Before Carter's talk, Hawking repeated that the „only *explanation*" of the isotropy of the universe was that founded on the suggestions of Dicke and Carter himself[150], while Wheeler (chairman on that day) pointed out that Hawking, Dicke „and Carter have touched such an interesting topic as the question whether man is involved in the design of the universe in a much more central way that one can previously imagine".[151]

It was indeed in some sense paradoxical that, in front of an audience gathered to address „tribute to the creator of the first scientific cosmological theory"[152], Carter remarked immediately that his intervention consisted: „basically of a reaction against exaggerated subservience to the *Copernican principle*".[153] Following Bondi, he intended here that criterion according to which we avoid to consider the Earth in a „central, specially favoured position"[154]; a criterion that, „unfortunately", has been sometime extended in a dogmatic manner.

Carter had in mind all those assumptions which invoke an extension of the usual cosmological principle to time in spite of Dicke's arguments and, in particular, that *perfect cosmological principle* which was assumed as a *dogma* in developing the steady-state model.[155] He underlined that the universe „is by no means homogeneous on a local scale" (say < 100 Mpc), aiming at last to pronounce (as he later wrote)[156]:

a warning to astrophysical and cosmological theorists of the risk of error in the interpretation of astronomical and cosmological information unless due account is taken of the biological restraints under which the information was acquired.

In other words, in 1973 Carter cautioned his audience to avoid the error of extending improperly any homology or uniformity assumption. One thing was to state that the Earth is not in a central position, another one was to conduct (consciously or subconsciously)[157] this statement to the extreme opposite, concluding that our position in space/time is absolutely typical and has nothing of peculiar or privileged.

---

As seen, the idea of a cyclical succession of many (possibly: infinitely many) closed universes, with *reprocessing* of the values of the fundamental constants and particle masses at the beginning of any new cycle, was anyway Wheeler's and not Carter's stuff.

[149] Carter, B.: Large number coincidences…, p. 291

[150] Hawking, S.W.: The anisotropy of the universe at large times, in: *Confrontation of cosmological theories with observational data*, Longair, M.S. (ed.), Dordrecht 1974, p. 283-286 (quotation from p. 285).

[151] Wheeler in: Longair, M.S. (ed.): *Confrontation* …, p. 287-288.

[152] From Zel'dovich's *Address* of Cracow's symposium, in: Longair, M.S. (ed.): *Confrontation* …, p. IX-XI.

[153] Carter, B.: Large Number Coincidences…, p. 291

[154] Bondi, H.: *Cosmology* … , p. 13. There is a letter of Carter to Don Page dated July 13th, 1994, where the author affirms explicitly that „the so called Copernican principle (that was implicitly used by Dirac in arguing for a theory of varying gravitational coupling …) postulates that our location in spacetime is entirely random a priori."

[155] Bondi, H.-Gold, T.: The steady-state theory of the expanding universe, *Monthly Notices of the Royal Astronomical Society* 108, 1948, p. 252-270.

[156] Carter, B.: The anthropic principle and its implications for biological evolution, *Philosophical Transactions of Royal Society* A310, 1983, p. 347-363, on p. 347.

[157] In 1973 Carter suggested that the *tendency* to extend the Copernican principle „to a most questionable dogma to the effect that our situation cannot be privileged in any sense" was „not always subconscious" (Carter, B.: Large number

The awareness that an abuse of the hypothesis of homology could be a source of bias, led Carter to formulate a methodological principle in order to avoid the formulation of cosmological principles based on an improper extrapolation of apparent symmetries.

The moral of all this in the case of the large adimensional numbers was that it is not necessary to elaborate unconventional extensions of physics and cosmology to explain the coincidences among them. In 1973, Carter re-named then the three categories of 1970 in accordance with what he now called the *anthropic principle*.

Category I coincidences were said to be of a „traditional kind", while category II coincidences became those which require the *weak anthropic principle* (WAP), i.e. a precept according to which[158]:

we must be prepared to take account of the fact that our location in the universe is *necessarily* privileged to the extent of being compatible with our existence as observers.

„More questionable" as they were[159], category III coincidences were said to require something more to be accepted as „complete physical explanations".[160] I.e.:
-      either an extension of the theory itself
-      or a new *philosophy* which makes possible to realize a combination of the ordinary WAP with an hypothesis on the existence of an ensemble of connected or disconnected branches of the universe over which „*fundamental constants* would have an extended range of values".[161]

The *strong anthropic principle* (SAP) was indeed a statement deriving from the „world ensemble philosophy"[162] and according to which[163] the universe (and hence the fundamental parameters on which it depends) must be such as to admit the creation of observers within it at some stage.

On this basis, Carter contemplated the eventuality of a systematical exploration of the constraints on the value of fundamental parameters that, at least in principle derive from the existence of living observers.

He admitted that SAP was not a „completely satisfying" criterion and that research for solutions based on a deeper „mathematical structure" was still preferable. At the same time, at least choosing such a suggestive terminology re-calling the *anthropos*, he surely advanced a proposal (if not, consciously or subconsciously, a provocation) destined to fuel interminable debates.

Some authors in fact soon read teleological (if not theological) implications in SAP, substituting (or sometime associating) a *taylor-made universe* for the many-universes philosophy.[164]

---

coincidences…, p. 291). In 1983 he wrote that the „extreme antithesis of the anthropocentric outlook was most dangerous as a source of biased thinking when it was adopted subconsciously" (Carter, B.: The anthropic principle and its implications …, p. 347).

[158] Carter, B.: Large number coincidences…, p. 293
[159] Ibid., p. 292
[160] Ibid., p. 295
[161] Carter, B.: The anthropic selection principle …, p. 54
[162] Carter, B.: Large number coincidences…, p. 298
[163] Ibid., p. 294. Immediately after this statement Carter inserted the famous passage: „To paraphrase Descartes, *cogito ergo mundus talis est*".
[164] Early teleological readings were advanced mainly after Wheeler's papers quoted in n. 56 & 131. With his ideas about a relevance of life and mind on the structure of the universe, Wheeler was surely influential on that part (not large, but indeed eminent) of the physics community that was involved with the anthropic reasoning. Apart from the already mentioned contributions, one may see also Wheeler's interview published on *Cosmic Search* 4, 1979 (nowadays available on www.bigear.org/vol1no4/wheeler.htm), where the author resumed Dicke's argument into the question „what good is a universe without somebody around to look at it?" and stated that „the anthropic principle looks at this universe, that universe and the other universe and rules out as mere meaningless machines all those in which awareness does not develop at some time".

Other sources that favoured a teleological reading of Carter's principle(s) were: Trimble, V.: Cosmology : Man's place in the universe, *American Scientist* 65, 1977, p. 77-86 ; Eccles, J.: *The human mystery*, New York 1979; Dyson, F.J.: *Disturbing the universe*, New York 1979 (A paper as Wald, G.: Fitness in the universe: choices and necessities, *Origins of Life* 5, 1974,

In effect, Carter's terminological choice recalled teleological overtones in more than a sense. For instance, just think to the fact that the term *anthropic* was coined by the Anglican theologian Frederick Robert Tennant who –at the end of the third decade of the XX century- aimed to a „wider teleology" in his *Philosophical Theology.[165]*

It is not my aim anyway to enquiry here the debate on the *anthropic reasoning* developed after Carter, or to discuss the anthropic principle(s) from an epistemological point of view.[166] I will limit myself to say that, during the 1980ies, Carter himself advanced an epistemological defence of his anthropic principle.[167] In a couple of occasions he then invoked a Bayesian approach to probability and stated again that scientists should be cautious on extending homology or symmetry assumptions. At last, Carter elaborated the following general statement:

Whenever one wishes to draw general conclusions from observations restricted to a small sample it is essential to know whether the sample should be considered to be biased, and if so how. The anthropic principle provides guidelines for taking account of the kind of bias that arises from the observer's own particular situation in the world.[168]

This statement confirms what the author affirmed clearly during the discussion that followed his talk at the Venice conference of 1988; i.e., that he presently regards the anthropic principle as a general principle of scientific enterprise rather than a cosmological one.

The 1980ies anyway brought with them also big developments in cosmology and fundamental physics. These induced theoretical physicists to accept the idea that certain features in the structure of the observable universe (e.g.: its size, its average density, the photon-to-baryon ratio) are particular exits of peculiar initial conditions, due to symmetry breaking phase transitions or to pseudo-casual processes that happened in the very early evolutionary history of the universe.[169] As a consequence, scenarios

---

p. 7-27, with all its references to a natural selection in the universe and to Henderson's *Fitness of the environment*, could appear as an important contribution in retrospect. Anyway it doesn't quote the anthropic principle).

The *resurgence in teleological views* in connection with the anthropic principle(s) reached a peak with the publication of the Barrow-Tipler essay (which is opened by a preface by Wheeler). The two authors advocated indeed an eutaxiological perspective (i.e.: aimed to point at the presence of a mathematical order and a „co-present, harmonius composition" of things in nature) rather than a teleological one. They outlined however a line of continuity between contemporary cosmology and a long series of philosophical speculations on the Design argument, and were surely responsible for large part of the common matching between anthropic reasoning and teleology.

To evaluate the impact of the teleological insights of Barrow-Tipler's essay, one may consult the following critical reviews: Press, W.H.: A place for teleology?, *Nature* 320, 1986, p. 315-316; Silk, J.: Teleological cosmology, *Science* 232, 1986, p. 1036; Gale, G: A revised Design: teleology and big questions in contemporary physics, *Biology and Philosophy* 2, 1987, p. 475-491; Craig, W.L.: Barrow and Tipler on the anthropic principle vs. Divine Design, *British Journal for the Philosophy of Science* 38, 1988; p. 389-395.

[165] Tennant, F.R.: *Philosophical theology*, two volumes, Cambridge 1928-1930. Cf. Barrow, J.D.-Tipler, F.J.: *The anthropic cosmological* …, section 3.9 and Craig, W.L.: *The teleological argument and the anthropic principle*, in: *The logic of rational theism: exploratory essays*, Craig, W.L.-McLeod, M. (eds.), Lewiston, N.Y. 1990, p. 127-153.

[166] Literature is full of papers on these topics, including some contributions of mine: Bettini, S.: *Il labirinto antropico, …* (quoted on n. 2 above); *The many faces of the anthropic principle*, forthcoming in *Memorie della Società Astronomica Italiana*.

[167] Carter, B.: *The anthropic principle: self-selection as …; The anthropic selection principle and the ultra-Darwinian synthesis,* in: *The anthropic principle, Proceedings of the second Venice conference on cosmology and philosophy November 1988*, Bertola, U.- Curi, V. (eds.), Cambridge 1993, p. 33-66. Amongst the wide literature on the epistemological status of the anthropic principle see for instance: Earman, J.: The SAP also rises: a critical examination of the anthropic principle, *American Philosophical Quarterly* 24, 1987, p. 307-317; Kanitscheider, B.: The anthropic principle and its epistemological status in modern physical cosmology, in: *Philosophy and the origin and evolution of the universe*, Agazzi, E.-Cordero, A. (eds.), Dordrecht 1991, p. 361-397; Bergia, S.: What, if anything, is the anthropic cosmological principle telling us?, in: Frontiers of fundamental physics, Barone, M.-Selleri, F. (eds.), New York, 1994, p. 73-82; Kirschenmann, P.P. 1994: Tautologie, methodologische Waarschuwing of Noodverklaring? een Kritische bespreking van enkele Antropische Principes, *Tijdschrift voor Filosofie* 56, p. 463-493, Rebaglia, A.: *Critica della ragione metascientifica. Argomenti antropici e spiegazioni scientifiche*, Milano 1996.

[168] This passage is taken by the draft of an article on the anthropic principle sent by Carter to Richard Matzner.

[169] While in the early 1980ies there were no compelling reasons to invoke seriously the concept of *many universes* in the form of disconnected domains, in the successive years that same concept was seen under a new light after a big change due

describing a multitude of causally disconnected branches of space/time called *bubbles*, *domains* or simply *universes* became customary in inflationary and quantum cosmology.

In recent times, moreover, while Carter has changed his mind with respect to both SAP and the many-worlds philosophy[170], a plethora of *anthropic principles* have been invoked in technical contexts and philosophical debates.

It seems that, although we surely have today a deeper understanding of various unsolved problems of the standard theory in comparison with the 1970ies, the advancement of knowledge has not made the need for anthropic reasoning obsolete. The emergence of many contemporary cosmological scenarios seems instead to have simply shifted anthropic *explanations* at a more profound theoretical level.

In conclusion I am then ready to affirm as a matter of fact that, although physicists have surely gone deeper in a series of issues since 1973, this does not seem to have eliminated at all the appeal to the anthropic reasoning.[171]

---

to the failure of early inflationary models, to new developments in fundamental physics and to the emergence (or persistence) of problems like those represented by the current value of the cosmological constant, the level of isotropy of the cosmic microwave background and the origin of cosmic structures. The cosmologists of the 1990ies moreover accepted the presence of a "random element in the initial evolutionary history of the universe" (Barrow, J.D.: Unprincipled cosmology, *Quarterly Journal of the Royal Astronomical Society* 34, 1993, p. 117-134, on p. 131) as a consequence of developments in fundamental physics.

During the 1980ies and the 1990ies, theories as Linde proposals (chaotic or eternal inflation) invoked then disconnected domains in the universe as a consequence of spontaneous symmetry breaking in very early phases and depicted at last the concept itself of a primeval explosion as a local event and our universe as a particular space/time domain of a globally stationary multiverse. All this contributed to make *many universes* scenarios a common occurrence.

[170] Cf. Carter, B.: The anthropic selection principle ..., p. 36. At the Venice conference of 1988 Carter confessed to find *antiquated* his old perspective on both SAP and the many worlds philosophy. He affirmed this explicitly in the (unpublished) discussion that followed Sciama's talk (which was indeed dedicated to the many universes). Anyway, Carter's new attitude emerged in Venice also from his own talk and from the discussion of Ellis, G.F.R.: *The anthropic principle: laws and environments* (in: Bertola, U.-Curi, V. (eds.): *The Anthropic principle*, ..., p. 27-36). He said: „it should be borne in mind that if the fine structure constant had been different one may guess that nature would have found alternative but not necessarily less effective mechanisms for doing the same jobs, so the argument for the strong anthropic principle (better described as the strong anthropic proposal) is not as strong as might at first appear".

[171] Among all the recent applications of the anthropic reasoning, I want to remember here just one example: Tegmark and Rees attempt to justify anthropically the value of the adimensional number $Q \equiv \Delta T/T$, which express the observed density fluctuations of the microwave background. See: Tegmark, M.-Rees, M.J.: Why is the cosmic microwave background fluctuation level $10^{-5}$?, *Astrophysical Journal* 499, 1998, p. 526-532.